**Lack of appropriate controls leads to mistaking *absence* seizures for post-traumatic epilepsy**

Krista M. Rodgers[1], F. Edward Dudek[2] and Daniel S. Barth[1]

[1]Department of Psychology and Neuroscience, University of Colorado, Boulder, CO 80309

[2]Department of Neurosurgery, University of Utah School of Medicine, Salt Lake City, UT 84108

**Introductory points.** We have read the comments by D'Ambrosio and colleagues, including their supplemental material; these comments have not changed our interpretation of our data (or theirs) or our conclusions. We stand fully behind our paper, as published. (1) We believe all aspects of our experimental design were completely appropriate for studying epileptogenesis; the experimental techniques for fluid percussion injury (FPI) have been used by us and others to study traumatic brain injury (TBI), and our experimental design has been used in other animal models to reveal genuine *bona fide* acquired epilepsy; (2) We agree that the data from the FPI-treated and control animals in our study were essentially identical, and by extension, we did not induce post-traumatic epilepsy; this was in fact a very important point of our paper; (3) The criteria for separating different types of epileptic seizures that we used is standard and well-accepted across the clinical and experimental literature. The claims of D'Ambrosio and colleagues refer to their prior attempts to re-define the features of different types of seizures to retroactively fit their data from their FPI experiments into a clinical context; integral to our concerns about their work is how they studied their control animals; and, (4) Although we did not cite some publications by D'Ambrosio and colleagues, we have cited extensive antecedent literature that is directly relevant to our data (and theirs); their experiments with anti-epileptic drugs and cooling are only indirectly relevant, and we believe these experiments also are questionable if not potentially flawed. Although SWDs appear to have been useful as models for the study of absence seizures, the idiosyncrasies of the sensitivity (or not) of EEEs to various AEDs is peripheral to the core issue. Reports that EEEs are pharmacoresistant to traditional AEDs and convenient models for testing possible new AEDs could reflect that EEEs are simply normal oscillations and not models of the complex partial seizures characteristic of PTE. In spite of the criticisms by D'Ambrosio and colleagues, we see no reason to change anything in our paper; our data directly contradict the hypotheses and conclusions of D'Ambrosio and colleagues, which we believe are fundamentally flawed for several reasons, particularly the widespread and consistent occurrence of epileptiform electrographic events (EEEs) in control animals.

**FPI technique.** D'Ambrosio et al. argue that we have not mastered the FPI technique to induce epileptogenesis and that our modified FPI using a Picospritzer (compared to the pendulum and piston device they and others use) has not been vetted for producing TBI sufficient to result in PTE. The FPI model using the pendulum and piston device produces a pressure wave that is highly sensitive to operational factors, and researchers have moved towards pneumatically driven FPI devices to improve reproducibility over older methods (Frey et al., 2009; Kabadi et al., 2010; Xiong et al., 2013). D'Ambrosio et al. report variability in pressure pulses between 8 - 10 ms, delivering between 3.0-4.0 atm across studies from 1999-2013. The pressure waveform generated by our FPI apparatus was found to be comparable to the morphology of pressure waveforms produced by pendulum and piston based devices, and also produced similar post-injury apnea, mortality, imaging (MRI), histopathology, and behavioral deficits (Frey et al., 2009). So, we submit that these methodological details are a distraction from the central theme of our paper. By using the rostral parasagittal fluid percussion injury (rpFPI) with 3 mm craniotomy and exactly the same injury parameters as D'Ambrosio et al. (and unique to their lab), we were unable to induce PTE seizures in any rats, nor has D'Ambrosio et al. The Picospritzer variant of the FPI technique, which is a published methods protocol (Frey et al., 2009) and was previously used by us (Rodgers et al., 2012, 2014), caused FPI with a macroscopically obvious injury, which was documented with histology (Frey et al., 2009; Rodgers et al., 2012, 2014). While the Picospritzer FPI injury has not yet been used as a model of epileptogenesis, the brain injuries most highly correlated with epilepsy after closed head injury (subdural hematomas and intracerebral contusions, Annegers et al., 1998) are induced by this injury device (see Fig. 3; Frey et al., 2009). That the FPI procedure was sufficient to cause other standard severe TBI outcomes in behavior is not in question. In their rebuttal, D'Ambrosio et al. describe the FPI technique - even with the device they used - as extremely difficult and highly variable, if not capricious and unreliable, in non-expert hands. The introduction of a Picospritzer instead of a pendulum and piston device was exactly because it delivers a far more accurate intensity and duration of impact under microprocessor control. We propose that it is not subtle details of method and skill set that makes all of the difference, as mentioned in our paper, it is that all others that have succeeded in producing *bona fide* PTE to our knowledge use a far more lateral and caudal impact location (above hippocampus) with a 5 mm craniotomy (effectively 3 times the dural surface area of impact) (Kharatishvili et al., 2006a, 2006b; Kharatishvili and Pitkanen, 2010; Shultz et al., 2013). These differences in procedure are not subtle and do not require special expertise.

**Electrographic Epileptiform Events (EEEs) of PTE are distinct from Spike-Wave Discharges (SWDs) of absence epilepsy.** D'Ambrosio et al. emphasize that we have only measured well-known SWDs of absence seizures and that the EEEs they report as post-traumatic seizures are distinct from SWD and clearly reflect PTE. Their distinction between SWD and EEE is based on four major points:

1) First, it is noted that the SWDs we and others measure start synchronously across multiple electrodes whereas EEE have a focal onset near the site of injury. It should be noted that this in the *only* distinction. The main quantitative point of our paper is that other than focal onset, EEEs are exactly like absence seizures on every other parameter shown in their publications. These include clear spike-and-wave morphology, high frequency of occurrence, short duration, lack of frequency progression during the event, lack of post-ictal suppression, "spindle-like" amplitude fluctuation with longer events, and time-locked behavioral interruption (i.e., "freezing") with accompanying facial automatisms. The only difference between EEEs and classic SWDs of absence epilepsy is identifiable focal onset near the site of injury. As we noted in our paper, the cortical focus theory even calls to question the concept of focal onset as a unique feature of absence seizures in rats (Meeren et al., 2002, 2005) or humans (van Luijtelaar et al., 2014). The argument of focal onset to distinguish EEEs does not reflect PTE in humans, which is not a focal variant of absence epilepsy, identical in every aspect to absence seizures except a focal versus generalized onset. For this reason we propose both here and in our paper that EEEs are not an etiologically realistic rat model of human PTE.

2) A second major point D'Ambrosio et al. make is that they have recognized and rejected idiopathic (absence/SWD) seizures in their results (See Fig. 2; D'Ambrosio et al., 2005) based on their consistently larger amplitude in occipital compared to frontal electrodes. All of our absence seizures were largest at frontal loci, either invalidating their discriminant criteria or indicating that our SWD events are not absence (idiopathic) seizures, both conclusions running counter to the main point of their rebuttal.

3) D'Ambrosio et al. argue that the SWDs we record in young adult (< 6 mo) rats are extremely rare and cast doubts on our methods for identifying SWD with supervised pattern recognition. This is an important point because, aside from focal onset of EEEs, the primary justification for why EEEs are distinct from SWDs is that they may be recorded in young rats. D'Ambrosio et al. cite the early age (<6 mo) of EEE onset as further evidence of its injury-induced origin compared to un-injured rats. Yet, to validate this point, they need to publish a quantitative study of control (un-injured or sham surgery) rats. Furthermore, their statement is not an accurate reflection of the literature on SWD. While more difficult to identify due to their short (1-2 sec) duration, numerous authors in several papers have seen SWDs in young as well as adult rats. For example, Pearce et al. (2014), in Scharfman's group, have reported that SWDs are present in 20% of young (2-3 months old) uninjured rats, a paper cited by the D'Ambrosio et al. rebuttal as evidence for their rarity in young animals. As we demonstrate, short bursts of what we term "larval" SWDs are detectable in rats as young as 1 mo and become longer and easier to quantify with advancing age. It is difficult to imagine quantifying brief SWD in the young animals with anything other than the supervised pattern recognition (support vector machine or SVM) we deployed. Trained on the SWD of each rat, the SVM model extracts hundreds of "candidate" SWDs, which are then visually verified for false positives. The increased prevalence, duration, and amplitude of SWDs with age in both injured and control animals gives the impression of epileptogenesis, but this age progression of SWD is a very common finding in most rat studies of absence epilepsy. If this is epileptogenesis, it is not injury induced since it exists equally in controls. SWD are in fact remarkably impervious to injury. D'Ambrosio et al. suggest that the use of young rats with a low background SWD would simplify the task of identifying EEE. We strongly recommend the opposite. EEEs should be induced in older rats (> 6 mo) with abundant SWDs so that the two phenomena can be quantitatively compared. This has never been done, or at least published. Discriminating EEEs and SWDs in the same rats is essential if EEEs are to be validated as a model of PTE.

4) D'Ambrosio et al cite evidence that other acquired epilepsy models report seizures that start with 7-9 Hz and, by comparison, prove that this is not a characteristic unique to SWD and that therefore their EEEs are not SWDs. As noted in more detail below, we emphasize that while SWDs are typically 7-9 Hz, spectral power is the *least* revealing parameter for identifying and comparing seizure activity. However, of more concern is that in their references to support their point, they mention – for example - work on a perinatal hypoxia model of acquired epilepsy (Rakhade et al., 2011) that has also adopted their EEEs as evidence for seizures, in this case even without focal onset (analysis of only a single electrode was shown). This is to be contrasted with a unilateral perinatal hypoxia-ischemia model studied with months-long continuous video-EEG monitoring; in this etiologically realistic model, a discrete brain infarct induced at postnatal day 7 leads to an initial preponderance of non-convulsive seizures that have an ipsilateral focal onset, last tens of seconds (up to minutes), and have a progressive evolution in waveform (Kadam et al., 2010). It concerns us that D'Ambrosio's EEEs have spread to influence other epilepsy models as well, all without a quantitative, age-dependent consideration of controls. And *controls* continue to be an essential issue. This issue has received no rigorous quantitative attention in the publications by D'Ambrosio's group. In recent papers, both FPI and sham-injured control rats are reused from previous studies, with a marked skew towards large numbers of FPI rats and small numbers of sham-injured controls 25/6 FPI/sham (D'Ambrosio et al., 2009) and 89/10 FPI/sham (Curia et al., 2011). The Rakhade et al. paper acknowledges 12.5% SWDs in their control (normoxic) Long Evans rats, which is actually surprisingly low given the reported 90% incidence in the same species at 4-6 mo age (Shaw, 2004). We contend that the spreading interpretation of EEEs as seizures of acquired epilepsy is alarming in light of repeated publication of evidence of SWDs in normal rats from a variety of outbred and inbred species (for review see: Pearce et al., 2014). The historical prevalence of SWD in outbred rats has justifiably raised questions concerning its interpretation even as absence seizures (Kaplan, 1985), but this is not the focus of the present letter.

In general, the suggestion that EEEs defined by D'Ambrosio et al. are in fact *not* distinct from SWDs and absence seizures is not new and was first raised critically by Kevin Kelly in direct response to another paper describing these events in normal animals of a different strain (Kelly, 2004a) and to the work by the D'Ambrosio lab (Kelly, 2004b). Furthermore, Kelly and co-workers published a paper on the photothrombotic stroke model in which they found no evidence for epilepsy, and only observed that events that were similar to those described by us and D'Ambrosio *in control animals* (Kelly et al., 2006).

**Human diagnostic criteria for genetic versus acquired epilepsy.** D'Ambrosio et al. state that we do not employ the correct diagnostic criteria used in humans to distinguish between the seizures of focal neocortical epilepsy and those of generalized absence epilepsy. This statement is inconsistent with the classic literature involving numerous clinical papers, only some of which we cited, that emphasize that brief (seconds long) events with a regular, non-evolving pattern, are characteristic of absence seizures, while longer lasting (i.e., tens of seconds, and up to minutes) electrographic seizures that progressively undergo an evolution in waveform are routinely expected for complex partial seizures (i.e., focal dyscognitive, Berg et al., 2010). These are long-standing criteria. For example, a classic 1993 video from the Epilepsy Foundation of America entitled "How to Recognize and Classify Seizures and Epilepsy" clearly distinguishes absence versus complex partial seizures on the basis of EEG morphology and duration. These criteria have been used extensively by many authors for decades, as cited in our paper. Thus, D'Ambrosio and colleague's statement that EEEs with an apparent focal onset are a clinically recognized criteria of PTE versus absence epilepsy is unsupported. Their statements about clinical relevance appear to relate to their own definition of what are clinically recognized criteria, similar to their proposed re-definition of "what is a seizure" (D'Ambrosio et al., 2009; 2010). This is not consistent with classical criteria for how to define absence and complex partial seizures. As noted earlier, PTE in humans is not a focal variant of absence epilepsy, identical in every aspect to absence seizures except a focal versus generalized onset.

Similar to the above point, D'Ambrosio et al. describe at some length the similarity of EEE to human focal seizures with spectral power in the theta frequency band (7-9 Hz). As noted, while SWD are typically 7-9 Hz, spectral power is perhaps the *least* revealing parameter for identifying and comparing seizure activity. The distinct quasi-periodic waveform of SWD (due to the spike and wave) does have a unique spectral signature with high values at a fixed fundamental frequency (7–9 Hz) and at whole multiples (harmonics) of this fundamental frequency, precisely the same as initially described by D'Ambrosio for EEEs (see Fig. 1 of (D'Ambrosio et al., 2004). This spectral signature is sufficiently different from simple theta activity (associated or not with focal seizures) that it has been used by some for SWD pattern recognition (Van Hese et al., 2003, 2009). We found characterization of SWD using autocorrelation functions in the time domain to be more reliable for this purpose. SWD is not theta activity and does not characterize focal seizures in rats or humans. Furthermore, D'Ambrosio argues that other seizure types, such as complex partial seizures, also have frequency components in 7-9 Hz range. We never said they did not. We only made the point that brief, regular, oscillatory-like activity - as observed in control and FPI-treated animals in our studies - had frequency components in the 7-9 Hz range similar to other models of SWDs observed as models of absence epilepsy. Thus, this argument is irrelevant because we agree that genuine seizures in models of acquired epilepsy – be they after status epilepticus, TBI, or stroke – do have spectral power at 7-9 Hz, but they also have many differences that make them distinct from the SWDs seen by D'Ambrosio and colleagues.

**Citation of previous literature.** Finally, D'Ambrosio et al. argue that we failed to cite or acknowledge published data from their group that are inconsistent with our conclusion; they cite three points. The first two points refer to work in their own lab (based on data with antiepileptic drugs and focal cooling) that represent additional evidence that EEEs in their laboratory are genuine epileptic seizures of PTE. Both of these arguments are completely dependent on *their own data* and do not represent independent corroboration of their work. The third point concerning spectral power in the theta band is addressed above and is essentially the same argument.

**Conclusions.** Thus, we disagree with all of the claims of D'Ambrosio et al. that our work is flawed, and we fully stand behind all aspects of the research, as described. Our purpose was to conduct anti-epileptogenesis studies in an animal model of PTE, which entailed replication of the studies of D'Ambrosio and colleagues; however, we were unable to replicate the critical data. All of the features of their claim of an animal model of PTE, aside from focal onset, were equally present in our control animals when the age and/or time-since-injury were comparable for the control and FPI-treated groups. We agree with their point that we did not observe any convulsive seizures. This may well be because of some idiosyncrasy of the FPI technique in his laboratory and in ours, since other groups have apparently observed convulsive seizures with similar FPI methods (Kharatishvili et al., 2006a, 2006b; Kharatishvili and Pitkanen, 2010; Shultz et al., 2013). The key result is that we observed events similar to numerous figures in their publications, *but they were also in control rats*. We did not observe any evidence of focal seizure onsets, but as reported by Kelly and others, occasional apparent focal onsets can occur with generalized epilepsies of the *absence* type. We caution other researchers to not trust that EEEs will only be seen after FPI; they are common in normal animals. Regardless of how faithfully one precisely follows the methods described by D'Ambrosio et al., one will eventually find EEEs in the control rats; these events are not unique to FPI-treated animals. If one looks carefully with appropriate analytical techniques, vestiges of EEEs or SWDs (i.e., "larval" EEE's) are detectable in young adults, but they are less frequent and shorter in duration, and most profound in older animals. These ages are in the range used by D'Ambrosio and colleagues.

We fully understand that D'Ambrosio and colleagues have provided a defensive response, since they created this model and have used it for over 13 years. What is unclear from their work is exactly when and how controls were done, and how the rats were paired and randomized. We emphasize again that our concerns are not new: Kevin Kelly and co-workers raised these concerns >10 ago (Kelly, 2004b), and other authors (Dudek and Bertram, 2010) raised them >5 years ago. Therefore, the observation in the Barth laboratory of these events in control animals prompted us to warn other researchers about potential problems with this model (and other models).

We expended three years and substantial DoD funds in an attempt to replicate the findings of D'Ambrosio and colleagues so we could go on to investigate PTE intervention strategies. We did indeed fail to produce PTE in any of the rats we studied. The objective of our report was to question whether the EEEs of D'Ambrosio et al. are reflective of PTE. Until appropriate controls are studied in relation to EEEs versus SWDs, we conclude that D'Ambrosio et al. have also failed to produce PTE. We are concerned that, since the first report by D'Ambrosio and colleagues more than a decade ago, no other laboratory has published a replication of EEEs as a model of PTE. We hope that this 10-year silence does not reflect lost time and funds on the part of others and that our report will at least serve as a warning to future investigators until this model of PTE has been rigorously vetted.

Conflict of Interest: F.E. Dudek has equity interest in and receives consultant fees from Epitel, Inc., which is a company that makes telemetric recording devices; however, this work did not use these devices. The other authors declare no competing financial interests.